\documentclass[aps,prl,twocolumn, showpacs]{revtex4}

\usepackage{epsfig}
\usepackage{graphicx}

\def\be{\begin{equation}}
\def\ee{\end{equation}}
\def\bea{\begin{eqnarray}}
\def\eea{\end{eqnarray}}

\def\XXint#1#2#3{{\setbox0=\hbox{$#1{#2#3}{\int}$}
     \vcenter{\hbox{$#2#3$}}\kern-.5\wd0}}

\begin{document}
\newcount\timehh  \newcount\timemm
\timehh=\time \divide\timehh by 60
\timemm=\time
\count255=\timehh\multiply\count255 by -60 \advance\timemm by \count255

\title{Spin-orbit coupling and spin transport}
\author{Emmanuel I. Rashba  }
\affiliation{Depatment of Physics, Harvard University, Cambridge, Massachusetts 02138, USA}
\date{\today}

\narrowtext

\begin{abstract}
Recent achievements in semiconductor spintronics are discussed. Special attention is paid to spin-orbit interaction, coupling of electron spins to external electric fields, and spin transport in media with spin-orbit coupling, including the mechanisms of spin-Hall effect. Importance of spin-transport parameters at spin-precession wave vector $k_{\rm so}$ is emphasized, and existence of an universal relation between spin currents and spin accumulation at the spatial scale of $\ell_{\rm so}\approx k_{\rm so}^{-1}$ is conjectured. 

\end{abstract}

\maketitle

\section{Introduction}
Spin is the only internal degree of freedom of electron. Employing it for creating electrical and optoelectronic devices with new functionalities is the ambitious goal of spintronics \cite{Wolf}. It is non less important that novel physics  including concepts that are exciting and challenging but sometimes controversial are involved. Last several years have witnessed remarkable experimental and theoretical achievements in this field.

Operating spintronic devices includes injecting nonequilibrium spins and manipulating spin polarization at given locations. Many of original proposals included electrical spin injection from magnetic (metallic or semiconductor) electrodes, similarly to spin injection into paramagnetic metals \cite{JohnSil}, and manipulating spin polarization by ac magnetic fields {\it via} Zeeman interaction. Modern trends in semiconductor spintronics are based on employing spin-orbit (SO) coupling for achieving both goals. It is believed, that such devices will provide efficient spin injection, and will be (i) controllable by electric fields allowing access to electron spins at nanometer scale, (ii) free from the parasitic effect of stray magnetic fields, and (iii) compatible with existing semiconductor technologies.

With the SO coupling as the central paradigm of semiconductor spintronics, developing a consistent theory of spin transport in systems with SO interaction is a demanding goal. The history of a related phenomenon, the anomalous Hall effect (AHE), indicates existence of fundamental theoretical problems. As different from the regular Hall effect, the AHE is driven by electron magnetization \mbox{\boldmath$M$} rather than by an external magnetic field \mbox{\boldmath$B$}, and therefore is based on SO coupling. The theory of AHE was initiated by Karplus and Luttinger back in 1954 \cite{KarLutt}, and 20 year long efforts resulted in the conclusion that the AHE is an extrinsic effect controlled by competing terms related to electron scattering. Some of the contributions to Hall current found in this way do not depend of the scattering time \cite{NozLew}, and recent research indicates existence of an intrinsic contribution to the anomalous Hall current that can be expressed in terms of Berry curvature in \mbox{\boldmath$k$}-space \cite{Jung,OnNa,Hald04}. Experiment still does not provide any convincing evidence regarding the mechanisms involved \cite{Lee04,Kats04}. Identification of spin transport mechanisms is even a more challenging task because the very notion of spin current is not well defined in media with SO coupling.

\section{Spin-orbit coupling}

In vacuum, SO coupling is described by the Thomas term in Pauli equation,
\begin{equation}
H_{\rm so}(\mbox{\boldmath$r$})=(e\hbar^2/4m_0^2c^2)~\mbox{\boldmath$\sigma$}\cdot(\nabla V(\mbox{\boldmath$r$})\times\mbox{\boldmath$k$}),
\label{eq1}
\end{equation}
that is small for nonrelativistic momenta $\hbar k\ll m_0c$, $V(\mbox{\boldmath$r$})$ being the scalar potential. In semiconductors with the gap $E_{\rm G}$ and SO band splitting $\Delta$ of a comparable magnitude, $E_{\rm G}\sim\Delta\sim 1$ eV, spin-orbit coupling is enhanced by a factor of about $m_0c^2/E_{\rm G}$. The specific form of the SO contribution to the free electron Hamiltonian is controlled by the symmetry requirements \cite{RS91}. Two typical SO Hamiltonians for electrons in three-dimensional (3D) crystals are
\begin{equation}
H_{\rm R}=\alpha(\mbox{\boldmath$\sigma$}\times\mbox{\boldmath$k$})\cdot{\hat{\bf z}},~
H_{\rm D}^{(3{\rm D})}=\beta_{3D}(\mbox{\boldmath$\sigma$}\cdot\mbox{\boldmath$K$}),
\label{eq2}
\end{equation}
where $K_z=k_z(k_x^2-k_y^2)$, and $K_x$ and $K_y$ can be obtained from $K_z$ by cyclic permutations. The term $H_{\rm R}$ (Rashba term) is typical of hexagonal A$_2$B$_6$ crystals. Being linear in \mbox{\boldmath$k$}, it is the leading term of the $\mbox{\boldmath$k$}\cdot\mbox{\boldmath$p$}$ - theory. Hence, it makes a profound effect on electron spin dynamics \cite{R60}. The term $H_{\rm D}^{(3{\rm D})}$ is typical of cubic A$_3$B$_5$ crystals and is known as the Dresselhaus 3D term.

In the Bloch function representation, SO coupling also changes the operator of coordinate $\mbox{\boldmath$r$}=i\nabla_{\mbox{\boldmath$k$}}+\mbox{\boldmath$r$}_{\rm so}$. For electrons in narrow-gap semiconductors, $\mbox{\boldmath$r$}_{\rm so}$ equals
\begin{equation}
\mbox{\boldmath$r$}_{\rm so}=(\hbar^2g/4m_0)[E_{\rm G}^{-1}+(E_{\rm G}+\Delta)^{-1}](\mbox{\boldmath$\sigma$}\times\mbox{\boldmath$k$})
\label{eq3}
\end{equation}
and is known as Yafet term \cite{Yaf}, $g$ being the $g$-factor. In modern terms, $\mbox{\boldmath$r$}_{\rm so}$ is Berry curvature in $\mbox{\boldmath$k$}$-space.

Spatial confinement lowers the system symmetry and makes appearance of \mbox{\boldmath$k$}-linear terms generic. In particular, $H_{\rm R}$ develops because and is controlled by the confinement asymmetry \cite{Nitta,Eng97} or external strain \cite{PiTi84} and is known as the structure induced asymmetry (SIA) or Rashba term, while $H_{\rm D}^{(3{\rm D})}$ reduces to the bulk induced asymmetry (BIA) or Dresselhaus term
\begin{equation}
H_{\rm D}=\beta(\mbox{\boldmath$\sigma$}_xk_x-\mbox{\boldmath$\sigma$}_yk_y),~~\beta\approx-\beta_{3{\rm D}}(\pi/d)^2,
\label{eq4}
\end{equation}
$d$ being the confinement layer width. For heavy holes, \mbox{\boldmath$k$}-linear terms are small and $k^3$-terms dominate,
\begin{equation}
H_{\rm hh}=\beta_{\rm hh}(k_+^3\sigma_--k_-^3\sigma_+),
\label{eq5}
\end{equation}
where $k_{\pm}=k_x\pm ik_y$ and $\sigma_\pm=\sigma_x\pm i\sigma_y$ \cite{Wink}. Relative magnitude of $\alpha$ and $\beta$ is material and geometry dependent. It was measured by electrical \cite{SOratio0,SOratio1}, optical \cite{Juss}, and photoelectrical \cite{GanPRL} means. Interesting effects of the interplay of $H_{\rm R}$ and $H_{\rm D}$ were predicted \cite{AlfBet1,AlfBet2,AGT05}.

The effect of SO coupling on spin transport critically depends on the analytical form of the SO Hamiltonian.

\section{Interplay of spin and orbit: \,\,\,\,\,\,\,\,\,\,\,\,\,\,\,\,\,\,\,\,\,\,\,\,\, Magnetic susceptibility}

Existence of SO interaction poses the questions (i) how deeply can this interaction influence macroscopic properties of a coupled system, and (ii) under which conditions spin dynamics can be separated (at least approximately) from the orbital dynamics. It is natural to begin with an equilibrium property like magnetic succeptibility $\chi$. For 2D systems, magnetization \mbox{\boldmath$M$} strongly oscillates as a function of \mbox{\boldmath$B$}, hence, the results are difficult to analyze \cite{BR84SS}. Explicit results are available for the dependence of low-\mbox{\boldmath$B$} susceptibility $\chi$ on the Fermi energy $\varepsilon_{\rm F}$ of 3D electrons at $T=0$ \cite{Boiko}. $H_{\rm R}$ electrons with  isotropic mass $m$ and $g$-factor in a magnetic field $\mbox{\boldmath$B$}\parallel{\hat{\bf z}}$ are paramagnetic for small $\varepsilon_{\rm F}$
\begin{equation}
\chi=\frac{e^2(\varepsilon_{\rm F}+\varepsilon_\alpha)}{8\pi\alpha mc^2}\biggl(1-\frac{mg}{2m_0}\biggr)^2,~~-\varepsilon_\alpha<\varepsilon_{\rm F}<0;
\label{eq6}
\end{equation}
here $\varepsilon_\alpha=m\alpha^2/\hbar^2$. The susceptibility $\chi$ vanishes when the Fermi level is at the spectrum bottom, $\varepsilon_{\rm F}=-\varepsilon_\alpha$, and increases linearly with $\epsilon_{\rm F}$. When the Fermi level passes through the singular point of the spectrum, $\varepsilon_{\rm F}=0$, the system becomes diamagnetic and $\chi$ diverges
\begin{equation}
\chi=-\frac{e^2\varepsilon_\alpha}{3\pi^2\hbar c^2\sqrt{2m\varepsilon_{\rm F}}},~~0<\varepsilon_{\rm F}\ll\varepsilon_\alpha.
\label{eq7}
\end{equation}
Further on, $\chi(\varepsilon_{\rm F})$ changes smoothly and asymptotically, for $\varepsilon_{\rm F}\gg\varepsilon_\alpha$, approaches the Landau-Pauli limit. 

Therefore, for $|\varepsilon_{\rm F}|\sim\varepsilon_\alpha$ spin and orbital motions are strongly coupled and their contributions cannot be separated. It is only when $\varepsilon_{\rm F}\gg\varepsilon_\alpha$ (and spin velocity $\alpha/\hbar\ll v_{\rm F}$) that the orbital motion becomes a perturbation to spin dynamics, and long Dyakonov-Perel \cite{DP72} spin relaxation times can be achieved \cite{KA2000}. For interference SO devices exploiting exact eigenstates, fulfillment of this criterion is not required.

\section{Basic ideas and experimental achievements}

A free-electron Hamiltonian with SO coupling $H_{\rm R}$ has, for each energy $\varepsilon$, two eigenstates with the momenta $k_\pm(\varepsilon)$ and opposite spin chirality. Datta and Das \cite{DD90} initiated the idea of spin transistor based on injecting spin polarized electrons and the following interference of the two components of the incident beam. The conductivity of the device depends on the phase difference $\Delta\Theta=(k_--k_+)L$ acquired by these states after crossing a ballistic sample of a length $L$ and oscillates with a period defined by the interference condition $(k_--k_+)L=2\pi n$, with $n$ an integer. An equivalent description of the same phenomenon is the precession of an electron spin in an effective magnetic field $\mbox{\boldmath$B$}_{\rm so}=(2\alpha/g\mu_{\rm B})(\mbox{\boldmath$k$}\times{\hat{\bf z}})$, $\mu_{\rm B}$ being the Bohr magneton. The device is controlled by the gate voltage $V_{\rm g}$ that modulates the SO coupling constant $\alpha$, $\alpha=\alpha(V_{\rm g})$. This seminal paper enjoyed a lot of attention and generated active discussions and new proposals. I am not aware of any practical realization of this device. A diffusive analog of it with spin coherence restored with a spatial period $L$ that does not depend on the driving electric field $E$ has been demonstrated recently \cite{CS05}; this periodicity proves that SO coupling was linear in $k$.

Koga {\it et al.} \cite{KNV04} proposed a ballistic spin-interferometer using square loop geometry with the interference of partial waves at the incident point. It does not require injecting spin-polarized electrons, is free of magnetic elements, and is completely controlled by the gate voltage $V_{\rm g}$. Remarkably, they achieved a phase difference $\Delta\Theta\sim\pi$ per loop for an extensive loop array; conductivity modulation was detected by the $V_{\rm g}$ dependence of Al'tshuler-Aronov-Spivak oscillations \cite{SKN05}. 

In an external magnetic field, spin polarized ballistic electron beams were created due to SO coupling in an open quantum dot serving as an emitter \cite{Folk} or in the bulk \cite{Rokhin}. For a spin-split energy spectrum, refraction and reflection of a spin-unpolarized electron beam transforms it into a system of spin-polarized beams \cite{Fin04,Ramag03}, and spin polarization of electrons scattered from a lithographic barrier has been achieved \cite{Chen05}.

Recently, experimental discovery of spin-Hall effect (SHE) with 3D electrons \cite{Kato04} and 2D holes \cite{Wund05} attracted attention. Dyakonov and Perel  \cite{DP71} predicted SHE due to Mott' skew scattering in a diffusive regime. It manifests itself in spin accumulation near the sample edges even in the absence of an external magnetic field. Because shortly before this discovery dissipationless spin currents for 3D holes in diamond type materials \cite{Mura03} and for 2D electrons in systems with SIA \cite{Sino04} have been proposed, the mechanism of SHE became a matter of active discussions. Below, I discuss this problem as applied to 2D systems.   

\section{Spin currents and magnetization}

A physically measurable quantity is magnetization, and when SO coupling is not too strong, $\varepsilon_{\rm F}\gg\varepsilon_\alpha$, one can distinguish orbital and spin contributions to it. It is appealing to discuss spin accumulation in terms of spin currents generated due to SO interaction and driven by an electric field $\mbox{\boldmath$E$}$; such an approach was employed by Governale et al. \cite{Gov03} and Mal'shukov et al. \cite{Mal04}. In vacuum, SO coupling exists only near scatterers and is described by Eq.~(\ref{eq1}). Scattered electrons are polarized along $(\mbox{\boldmath$k$}\times\mbox{\boldmath$k$}')$, where $\mbox{\boldmath$k$}$ and $\mbox{\boldmath$k$}'$ are the initial and final momenta, respectively. Electrons scattered to the left and right  have opposite polarizations. Spin currents are conserved in the free space.

In semiconductors, in addition to this extrinsic effect, there exists also an intrinsic SO effect originating due to the spin dependence of the free electron Hamiltonian $H(\mbox{\boldmath$k$})$. Calculating the cumulative effect of both mechanisms to spin transport in terms of spin currents is problematic for several reasons.

(i) Because of the torque produced by the field $\mbox{\boldmath$B$}_{\rm so}$, spin is not conserved and does not obey the continuity equation. Therefore, the standard procedure cannot be applied for defining spin currents. Usually they are defined similarly to the media without SO coupling as 
\begin{equation}
J_{{\rm s},i}^j =\frac{1}{2}\langle v_i\sigma_j+\sigma_jv_i\rangle,~~i=\{x,y\},~~j=\{x,y,z\}, 
\label{eq8}
\end{equation}
where the velocity $\mbox{\boldmath$v$}$ is defined, with the due account of the SO coupling in the bulk, as $\mbox{\boldmath$v$}=\hbar^{-1}\partial H(\mbox{\boldmath$k$})/\partial\mbox{\boldmath$k$}$. Competing definitions of $J_{{\rm s},i}^j$ were also proposed \cite{Niu}.

(ii) Spin currents of Eq.~(\ref{eq8}) are even with respect to time inversion, $t\rightarrow -t$, hence, some of the components of $J_{{\rm s},i}^j$ may be nonzero in equilibrium. E.g., for the Hamiltonian $H_{\rm R}$, $J_{{\rm s},x}^y=-J_{{\rm s},y}^x\neq0$. In this respect, spin currents are reminiscent of the gas pressure that is the momentum flux. Similarly to it, the currents $J_{{\rm s},i}^j$, depending on the conditions, can play a role of both thermodynamic and transport parameters \cite{R05}.

(iii) Because in equilibrium $J_{{\rm s},i}^j\neq0$ in media with SO coupling, and $J_{{\rm s},i}^j=0$ in media without it, spin currents are not conserved at their interfaces.

(iv) $J_{{\rm s},i}^j$ are $t$-even while the magnetization $\mbox{\boldmath$M$}$ is $t$-odd, hence,  magnetization build-up near a boundary should involve some $t$-inversion breaking process.

Because of these problems, bulk spin currents of Eq.~(\ref{eq8}) seem not be directly related to spin accumulation near edges. Nevertheless, calculations of them turned to be highly instructive regarding the dependence of spin transport on SO coupling mechanisms. 

\section{Effect of electron scattering on spin currents }

Dissipationless spin currents were found originally by applying the Kubo formula in the regime when scattering was disregarded. Such an approach is questionable because the linear response provided by this formula can be maintained only due to scattering. A proper account for electron scattering resulted in different results for the Hamiltonians $H_{\rm R}$ and $H_{\rm hh}$, and the Luttinger Hamiltonian of 3D holes. For $H_{\rm R}$ electrons, Inoue et al. \cite{Inoue}, followed a number of different authors \cite{MSH04,RS05,Dimit}, have found dramatic cancelation of the bubble and ladder diagrams for $J_{{\rm s},x}^z$, hence, spin current vanishes, and this is true for any type of elastic scattering \cite{Kha}. On the contrary, for more general SO coupling mechanisms, $J_{{\rm s},x}^z\neq0$ \cite{Murak04,MC05,BZ04}. In particular, for the Hamiltonian $H_{\rm hh}$ ladder diagrams vanish \cite{BZ04}, and $J_{{\rm s},x}^z$ is described by bubble diagrams \cite{SL05}.

Dimitrova \cite{Dimit} related spin-current cancelation for the operator $H_{\rm R}$ to the operator identities 
\begin{equation}
{\dot\sigma}_x=-k_\alpha{\hat J}_{{\rm s},x}^z,~~{\dot\sigma}_y=-k_\alpha{\hat J}_{{\rm s},y}^z,~~k_\alpha=\alpha m/\hbar^2, 
\label{eq9}
\end{equation}
$k_\alpha$ being the spin-precession wave vector. Eq.~(\ref{eq9}) implies that non-zero macroscopic spin currents, $J_{{\rm s},x}^z\neq0$, $J_{{\rm s},y}^z\neq0$, would be tantamount to a time dependent macroscopic magnetization. This argument is valid for a generic combination of $H_{\rm R}$ and $H_{\rm D}$ ($\alpha\neq\pm\beta$); for more detail see Refs.~\cite{CL05} and \cite{Bauer05}. It also persists for spin-independent electron scattering.

A different insight into spin-current cancelation comes from applying the Kubo formula to $H_{\rm R}$ electrons in a magnetic field $\mbox{\boldmath$B$}\parallel{\hat{\bf z}}$ \cite{R04}. This system is exactly soluble \cite{R60}, and for calculating transverse responses introducing impurity scattering is not required. Spin current $J_{{\rm s},x}^z$ includes two contributions. First comes from interbranch transitions  and includes a sum over the Fermi sea region  with $k_+<k<k_-$. Second comes from two intrabranch transitions at the Fermi sea edges, one per branch. For $g=0$, these terms cancel for arbitrary $B$ and $\alpha/\hbar v_F$ because of exact sum rules specific for $H_{\rm R}$ Hamiltonian, hence, $J_{{\rm s},x}^z=0$. They ensure insensitivity of $J_{{\rm s},x}^z$ to scattering mechanisms.

For $B\rightarrow0$, the separation $\hbar\omega_c$ between adjacent levels vanishes, $\omega_c$ being the cyclotron frequency. Thence, the intrabranch contribution to spin conductivity is highly sensitive to any inaccuracies in numerical simulations, while the interbranch contribution equal to the universal conductivity $\sigma_{\rm sH}=e/4\pi\hbar$ by Sinova et al. \cite{Sino04} is robust. That is why special efforts were needed \cite{Sheng05,Nom05} to bring numerical results on $H_{\rm R}$ Hamiltonian in agreement with the analytical ones.

\section{SHE: frequency and momentum dependence}

Sensitivity of intrabranch terms to the external parameters manifests itself in a strong frequency dependence at the scale of $\omega\sim\omega_c$ \cite{R04}. With $\omega\rightarrow i/\tau$, where $\tau$ is the momentum relaxation time, $\sigma_{\rm sH}(\omega)$ matches the result of Ref.~\cite{MSH04}. In the frequency range $\omega_c\ll\omega\ll2\alpha k_\pm/\hbar$, spin conductivity $\sigma_{\rm sH}(\omega)\approx e/4\pi\hbar$.

Similarly, $\sigma_{\rm sH}$ shows momentum dependence at the scale of $k\sim k_\alpha$ \cite{R05}. To find spatial dependence of spin accumulation $\mbox{\boldmath$S$}(\mbox{\boldmath$r$})$, we take advantage of using diffusive equations derived by Mishchenko et al. \cite{MSH04} and Burkov et al. \cite{BNMD04}. For $H_{\rm R}$ electrons in the $x\geq0$ half-plane driven by a field $\mbox{\boldmath$E$}\parallel{\hat{\bf y}}$,
\begin{eqnarray}
(DS''_x-S_x/\tau_{\rm s})+2k_\alpha\ell v_FS'_z&=&S_x^\infty/\tau_{\rm s},\nonumber\\
(DS''_z-2S_z/\tau_{\rm s})-2k_\alpha\ell v_FS'_x&=&0,\,\,S_y=0, 
\label{eq10}
\end{eqnarray}
where $\ell=v_F\tau$, $D=v_F^2\tau/2$, $\tau_{\rm s}=1/2\tau(k_\alpha v_F)^2$, and $S_x^\infty=k_\alpha\tau eE/2\pi$ are the mean free path, diffusion coefficient, spin relaxation time, and spin accumulation at $x\rightarrow\infty$, respectively. $S_x^\infty$ has been measured recently \cite{KatoSO,GanSO}. Eq.~(\ref{eq10}) is valid when $k_\alpha\ell\ll1$. The boundary conditions for it should relate $\mbox{\boldmath$S$}$ and $\mbox{\boldmath$S$}'$ at $x=0$. 

\begin{figure}[h]
\begin{center}\leavevmode
\includegraphics[width=0.7\linewidth]{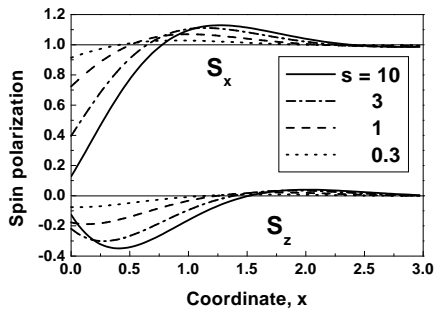}
\caption{Spin accumulation of $H_{\rm R}$ electrons in the $x>0$ half-plane driven by a field $\mbox{\boldmath$E$}\parallel{\hat{\bf y}}$. Upper curves: $S_x$, lower curves: $S_z$; all in units of $S_x^\infty$. Dimensionless spin relaxation velocity ${\rm s}=s/k_\alpha D$, the coordinate $x$ in units of $k_\alpha^{-1}$.}
\label{figurename}\end{center}\end{figure}

In Fig.~1 are shown the results for the boundary condition $DS'_i(0)=sS_i(0)$, $i=x,y$, where $s$ is the surface spin-relaxation velocity. The condition $s\approx0$ sounds reasonable for smooth lithographic barriers along free edges, hence, we corroborate the conclusion by Mishchenko et al. \cite{MSH04} regarding the absence of spin accumulation along them. By engineering high $s$ edges, one can achieve spin accumulation along them. All curves in Fig.~1 are scaled in units of $S_x^\infty$, hence, the process looks like rotation of $\mbox{\boldmath$S$}$ from the $\hat{\bf x}$ to $\hat {\bf z}$ direction due to the transverse spin flow. Leaking of spins into $\alpha=0$ reservoirs results in a similar effect \cite{Bauer05}.

SHE due to spin-current cancelation breaking should also develop in systems with non-planar boundaries and inhomogeneities. Stern-Gerlach spin filter due to $\alpha=\alpha(x)$ was proposed by Ohe et al. \cite{Ohe}.

Remarkably, SHE of Fig.~1 has been found for $H_{\rm R}$ electrons, i.e., for a system with $J_{{\rm s},x}^z=0$. This fact indicates that bulk currents $J_{{\rm s},i}^j$ are not enough for calculating SHE, and transport equations valid at the spatial scale of $\ell_\alpha=k_\alpha^{-1}$ are needed. For $H_{\rm hh}$ holes they are not available yet, hence, consistent treatment of the data by Wunderlich et al. \cite{Wund05} is not possible. However, if one accepts that transport spin currents at the spin precession wave vector $k_{\rm so}=k_\beta\sim m\beta_{\rm hh}k_{\rm F}^2/\hbar^2$ are governed by  ``macroscopic" spin conductivity and are about $eE/\hbar$ \cite{BZ04}, one can speculate about SHE. Indeed, $S_z(0)\sim\hbar J_{{\rm s},x}^z/v_{\rm eff}$, and $v_{\rm eff}\sim k_\beta^{-1}/\tau$ is a proper estimate for spin-transport velocity. Hence, $S_z(0)\sim k_\beta\tau eE$, similarly to $S_x^\infty$ found above, and large SHE of Ref.~\cite{Wund05} can be attributed to large magnitude of $k_\beta\tau$.

Generalizing the above conclusions, we make a {\it Conjecture} that, irrespective to specific SO coupling mechanisms, spin currents at the wave vector $k_{\rm so}$ have magnitudes about the {\it ``universal"} value of $eE/\hbar$ \cite{Sino04} and acquire the meaning of {\it transport} currents. Hence
\begin{equation}
J_{\rm sH}(k_{\rm so})\sim eE/\hbar\,,\,\, S(k_{\rm so})/\hbar\sim k_{\rm so}\tau eE/\hbar\,. 
\label{eq11}
\end{equation}
Numerical coefficients in these equations depend on the specific form of $H_{\rm so}$ and on boundary conditions.

Difference in SHE in meso- and macroscopic $H_{\rm R}$ conductors, emphasized in Ref.~\cite{NicolMeso}, seems to come from the momentum dependence of spin conductivity. It also enables spin injection from quantum rings \cite{Kis03,Niko05}.

\section{Extrinsic SHE in {\it n}-GaAs}

Engel, Halperin and the present author \cite{EHR05} have developed a theory of extrinsic SHE in $n$-GaAs. Its magnitude comes from the enhancement of the coefficient in Eq.~(\ref{eq1}) by a factor of more than $10^6$ due to the coupling across the gap $E_{\rm G}$. A Boltzmann theory of SHE results in skew scattering and side jump terms that are of comparable magnitude for materials with small $k_{\rm F}\ell\approx 4$ used in Ref.~\cite{Kato04}. Remarkably, the side jump term includes a Berry phase part that might indicate a way for unifying the intrinsic and extrinsic approaches to SHE. The results are in a reasonable agreement with the data by Kato et al. \cite{Kato04}. More recently, Tse and Das Sarma came to similar conclusions \cite{TDS}.

\section{Acknowledgments}

I am highly grateful to H.-A. Engel and B.I. Halperin for close collaboration and inspiring discussions. Financial support from Harvard Center for Nanoscale Systems and DARPA is acknowledged.

\end{document}